\documentclass[a4paper,fleqn,useAMS,usenatbib]{mnras}

\pdfoutput=1

\usepackage{mathptmx}
\usepackage[T1]{fontenc}
\usepackage{ae,aecompl}

\usepackage{geometry}
\usepackage{pdflscape}

\usepackage[dvips]{graphicx}
\usepackage{amsmath}
\usepackage{amsfonts}
\usepackage{amssymb}
\usepackage{comment}
\usepackage{wasysym}
\usepackage[dvipsnames]{xcolor}
\usepackage[%
        ]{hyperref}
        
\hypersetup{
	colorlinks=true,	% false: boxed links; true: colored links
	urlcolor=MidnightBlue,		% color of external links
	pdfpagelabels=true,
	hypertexnames=true,
	plainpages=false,
	naturalnames=true,
	pdftitle={Fast TTVs},    % title
	pdfauthor={Kipping and Yahalomi},     % author
	linkcolor=WildStrawberry,          % color of internal links (change box color with linkbordercolor)
	citecolor=ForestGreen,        % color of links to bibliography
}

\newcommand{\kepler}{{\it Kepler}}
\newcommand{\gaia}{{\it Gaia}}

\newcommand{\Pttv}{P_{\mathrm{TTV}}}

\newcommand{\isochrones}{{\tt isochrones}}
\newcommand{\luna}{{\tt LUNA}}
\newcommand{\SpecMatch}{{\tt SpecMatch}}
\newcommand{\multi}{{\tt MultiNest}}
\newcommand{\cofiam}{{\tt cofiam}}
\newcommand{\polyam}{{\tt polyam}}
\newcommand{\local}{{\tt local}}
\newcommand{\gp}{{\tt gp}}

\newcommand{\forecaster}{{\tt forecaster}}
\newcommand{\wwwcoolworlds}{\href{https://github.com/CoolWorlds/fastTTVs}{this URL}}

\title[A search for TTVs within the exomoon corridor]{
A search for transit timing variations within the exomoon corridor using \kepler\ data
}
\author[Kipping \& Yahalomi]{David Kipping$^{1}$\thanks{E-mail:
\href{mailto:dkipping@astro.columbia.edu}{dkipping@astro.columbia.edu}} and Daniel A.. Yahalomi$^{1}$ \\
$^{1}$Dept. of Astronomy, Columbia University, 550 W 120th Street, New York NY 10027}

\date{Accepted 2022 November 10. Received 2022 November 10; in original form 2022 August 31}

\pubyear{2022}

\begin{document}
\label{firstpage}
\pagerange{\pageref{firstpage}--\pageref{lastpage}}
\maketitle

\begin{abstract}
An exomoon will produce transit timing variations (TTVs) upon the parent planet
and their undersampled nature causes half of such TTVs to manifest within a
frequency range of 2 to 4 cycles, irrespective of exomoon demographics. Here,
we search through published \kepler\ TTV data for such
signals, applying a battery of significance and robustness checks, plus
independent light curve analyses for candidate signals. Using the original
transit times, we identify 11 (ostensibly) single-planets with a robust,
significant and fast ($P_{\mathrm{TTV}}<4$ cycles) TTV signal. However, of
these, only 5 are recovered in an independent analysis of the original
photometry, underscoring the importance of such checks. The surviving signals
are subjected to an additional trifecta of statistical tests to ensure signal
significance, predictive capability and consistency with an exomoon.
KOI-3678.01, previously validated as Kepler-1513b, is the only case that passes
every test, exhibiting a highly significant ($>20\sigma$) TTV signal with a
periodicity, amplitude and shape consistent with that caused by an exomoon. Our
analysis finds that this planet is $8.2_{-0.5}^{+0.7}$\,$R_{\oplus}$ orbiting
at $0.53_{-0.03}^{+0.04}$\,AU around a late G-type dwarf. After forecasting the
planetary mass, we expect it to be capable of maintaining at least a
0.3\,$M_{\oplus}$ exomoon for 5\,Gyr, and the TTV signal corresponds to a moon
mass as low as $0.75$ Lunar masses. We thus encourage follow-up
observations and dynamical analysis of this unique signal, but caution skepticism
until such data can be obtained.
\end{abstract}

\begin{keywords}
planets and satellites: detection --- methods: data analysis --- techniques: photometric
\end{keywords}

\section{Introduction}

The Solar System reveals that moons appear to be a natural by-product of planet
formation. Indeed, at least three distinct mechanisms appear required to explain
the largest moons ($0.2$-$0.4$\,$R_{\oplus}$); a capture mechanism for Triton
(e.g. see \citealt{agnor:2006}), a giant impact for the Moon
\citep{reginald:1946} and in-situ formation disk (e.g. see
\citealt{canup:2002}). If transiting exoplanets possess moons, then they
should be expected to also transit, as well as perturb the path of the parent
planet such as via TTVs, transit timing variations \citep{sartoretti:1999}.
Whilst the detection of both effects is ultimately sought, yielding a radius
and mass respectively, the two effects will in general have different
signal-to-noise ratios (SNRs), meaning that for observations at the threshold
of detectability, we should only expect one to be initially found.

To date, there is only one known example where both signatures appear to
manifest - Kepler-1625b-i \citep{teachey:2018}. Although the TTV signal here
appears robust, the moon transit was independently recovered by one
team \citep{heller:2019} but not another\footnote{However, a comparative
study of the different light curve reductions indicated that the reduction
of \citet{kreidberg:2019} appeared to be affected by systematics.}
\citep{kreidberg:2019} - thus leaving the situation ambiguous without
follow-up observations. Extensive searches for additional moon-like transits
have been attempted (e.g. see \citealt{HEK:2012} and subsequent papers in that
series) but only one other \kepler\ candidate has been reported to exhibit such
a signature, the exomoon candidate Kepler-1708b-i \citep{k1708b:2022}. In that
case the TTV is presently undetectable since only two epochs have been observed
to date.

Despite the considerable focus on moon transits, TTVs were amongst the first
methods proposed to look for exomoons \citep{sartoretti:1999}, have been
cataloged for hundreds of systems (e.g. \citealt{mazeh:2013}) and dozens
of confirmed TTV systems now exist thanks to their planet-planet interactions
(e.g. \citealt{hadden:2014}). The classic problem with TTVs is indeed this
latter point though, how can one discern whether a given TTV signal is
due to an exomoon or a perturbing planet? Transit duration variations (TDVs)
can break this degeneracy in ideal conditions \citep{kipping:2009a}, but in
practice TDV measurements are rarely precise enough to tease out these effects.
Radial velocity (RV) measurements could also come to the rescue, by eliminating
the hidden planet hypothesis, but here too sufficiently precise RV measurements
of the faint \kepler\ stars have been lacking for the majority of planets. One
recently highlighted ``statistical'' answer is to look at the TTV frequency.

This idea was described in \citet{corridor:2021}, where it was shown that
exomoon induced TTVs will always be undersampled and thus their signatures will
manifest as aliases. These aliases are much more likely to occur close to
the Nyquist frequency and in fact 50\% of all of these aliases will occur
in a period range of 2-4\,cycles; the so-called ``exomoon corridor''. What's
remarkable is that this result is independent of the exomoon population
properties, and is distinctive from the typically much longer periodicities
planet-planet interactions produce \citet{hadden:2014}.

Pulling on this thread, we here conduct a search for TTV signals in the
exomoon corridor. We start with the \citet{holczer:2016} catalog
(H+16 hereafter), derived from the \kepler\ data, as our initial input, which
is then filtered down to the most exomoon-like signatures in
Section~\ref{sec:filtering}. We then independently analyse the light curves of
the best candidates to confirm their candidacy in Section~\ref{sec:indep}. We
discuss our only surviving candidate in Section~\ref{sec:3678} before 
highlighting broader implications of this work in Section~\ref{sec:discussion}.

\section{Filtering of the H+16 Catalog}
\label{sec:filtering}

Our analysis makes use of the H+16 transit timing catalog derived from the
\textit{Kepler Mission}. Despite being published six years ago, H+16 remains the
most up-to-date dedicated public \kepler\ TTV catalog at the time of writing.
H+16 report that on 23rd November 2013, 4690 KOIs were listed in the NASA
Exoplanet Archive after eliminating false positives, whereas on 19th July 2022
that number is marginally higher at 4716. A large number of KOIs were deemed
unsuitable for TTV analysis by H+16, for having insignificant depths,
excessively large depths, periods greater than 300\,days, or for being a known
false alarm/eclipsing binary. After these cuts, 2599 KOIs were used to produce
the H+16 TTV catalog.

We note that many of the KOIs that H+16 ignored can be filled in using the
transit times available from \citet{rowe:2014} and \citet{rowe:2015}. However,
by virtue of H+16's filters, these are inherently lower quality KOIs to work
with, and would also lead to a non-homogeneous input catalog. For these reasons,
we elect to not include them here.

H+16 produced a summary list of KOIs which they concluded to have a significant
($p$-value $<10^{-4}$), long-term ($\Pttv>100$\,days) TTV signal (their
Table~5), and also another list of those with a significant, short-term
($3<\Pttv<80$\,days) TTV signal (their Table~7). A scoring system that rests
primarily on $p$-values is somewhat precarious as it truly only ranks the
``surprisingness'' of an event, and is commonly misinterpreted to equate to the
probability that the null hypothesis (i.e. ``there is no TTV'') is true
\citep{colquhoun:2014,wasserstein:2016}. Certainly such cases are deserving of
further attention, but in isolation an extreme $p$-value leaves room for
ambiguity about the reality of the putative signal. Further, the
short/long-term definitions used by H+16 are framed in terms of absolute
temporal units (i.e. $\Pttv$), whereas the exomoon corridor we seek is defined
on a relative temporal scale (i.e. $\Pttv/P$).

We also note that \citet{kane:2019} produced an independent visual ranking of
TTV systems, leveraging the H+16 catalog as well as that of \citet{rowe:2014}
and \citet{rowe:2015}. Whilst we will compare our results to earlier work at
the end, we elected to conduct our own selection procedure for significant TTVs
that focusses more on inferential statistical measures, and only utilizes
$p$-values for flagging \textit{potentially} spurious cases, rather than
flagging candidate TTV detections.

\subsection{Lomb-Scargle periodograms}
\label{sub:LSperiodogram}

In searching for an exomoon corridor signal, we highlight two key features that
affect our strategy: i) an exomoon TTV signal is expected to be strictly
periodic \citep{sartoretti:1999,kipping:2009a}, and ii) the transit timing
catalog from H+16 is often sparse. Accordingly, periodic signals should be
sought using a Lomb-Scargle (LS) periodogram \citep{lomb:1976,scargle:1982}.

Crucially, we highlight that the LS periodogram is ideally run on the transit
times, not the TTVs. By definition, TTVs are the transit times with a linear
ephemeris subtracted, and that linear ephemeris must have been derived from
a regression of effectively the transit times (at least with this data set).
In principle, there's no reason why one can't indeed do two regressions of
the transit times; one that derives the linear ephemeris, and then a
separate second one that fits a periodic signal through the TTVs. However,
such a two-stage process does not trivially propagate the uncertainty of
the linear ephemeris itself into the inference of the periodic
signal\footnote{Indeed said error propagation is not conducted in the analysis
of H+16.}. This could be accommodated by utilising the covariance function
derived from the linear ephemeris, but it's far simpler and more robust to just
fit the transit times directly to a linear ephemeris plus sinusoidal model.

Accordingly, our LS periodogram defines a grid of trial periods and then
at each period regresses two models to the H+16 transit times. The null
model is simply:

\begin{align}
\tau_{\mathrm{null},e} &= P_{\mathrm{null}}*e + \tau_{\mathrm{null},\mathrm{ref}},
\end{align}

where $e$ is the epoch number, $P$ is the planetary period,
$\tau$ is the time of transit minimum and the subscripts ``null''
and ``ref'' refer to the null hypothesis model and the reference
epoch respectively. The second model we regress is that of a linear
ephemeris with a sinusoidal variation added on top:

\begin{align}
\tau_{\mathrm{sin},e} &= P_{\mathrm{sin}}*e + \tau_{\mathrm{sin},\mathrm{ref}} + a_S \sin(n_{\mathrm{TTV}}e) + a_C \cos(n_{\mathrm{TTV}}e),
\end{align}

where $n_{\mathrm{TTV}}$ is the trial frequency ($\equiv 2\pi/\Pttv$), and
$a_S$ and $a_C$ are the two linear amplitude components of a sinusoidal wave.
By writing the model out this way, and working along a pre-defined frequency
grid, the problem is linear with respect to the unknown parameters and thus
we exploit a linear solver to infer the best-fitting model.

Outlier transit times can significantly distort the results of an
LS periodogram, which is predicated upon mean-based (weighted linear least
squares) statistics. To alleviate this somewhat, we apply a simple outlier
rejection scheme to the H+16 transit times. To this end, we first extract the
quoted TTVs from H+16 and divide them by their quoted uncertainties. We next
measure the RMS of this list using a median-based robust measure, specifically
1.4826 multiplied by the median absolute deviation (MAD). We then remove any
transit times for which the quoted TTV normalised by the quoted uncertainty
exceeds 10 times this robust RMS value. The idea here is to reject points which
are dispersed an order-of-magnitude more than the observed scatter. We also
remove any transit times for which the H+16 reported uncertainty is $>3$ times
greater than the median uncertainty of that KOI, typically associated
with partial transits of poorer data quality.

For the period grid, we define the shortest period using the Nyquist rate
\citep{nyquist:1928}, which is given by the twice the minimum temporal spacing
between any two transit times (almost always 2 cycles). The longest period is
given by twice the temporal baseline of the transit times. In principle, the
number of periods/frequencies scanned should be equal to the number of transit
times \citep{vanderplas:2018}, but in practice we overscan by a factor of ten
to create a smooth, dense periodogram. At each period, we evaluate the
$\chi^2$ of two competing models, which are then appended and saved to disk.
We also save the amplitude of the best fitting TTV signal
($=\sqrt{a_S^2+a_C^2}$), the number of available transit times, and finally
a modification of the $\chi^2$ metric using median statistics that we
dub $\Xi^2 = n\,\mathrm{median}[(\mathbf{r}/\boldsymbol{\sigma})^2]$.

After this effort, we are now ready to apply our first filter to the H+16
catalog. Specifically, we cut down to KOIs for which the peak $\Pttv$
(as derived from our LS periodogram) is less than 20\,cycles (the $\simeq$95\%
confidence limit of the exomoon corridor; \citealt{corridor:2021}) \textit{and}
the Bayesian Information Criterion (BIC; \citealt{schwarz:1978}) favours the
sinusoidal model. This first filter can be thought of as a ``rough cut'' for 
potentially interesting exomoon corridor signals. The filter reduces the
number of KOIs from 2599 to 1822. Thus, taking the H+16 transit times at
face value, there is a very large number of KOIs which seemingly exhibit
evidence (or at least ``a statistical preference'') for periodic TTVs
\footnote{A similar result was also reported by \citet{impossible:2020}.}.

\subsection{Remove false positives and multis}

\begin{figure*}
\begin{center}
\includegraphics[width=17.4cm,angle=0,clip=true]{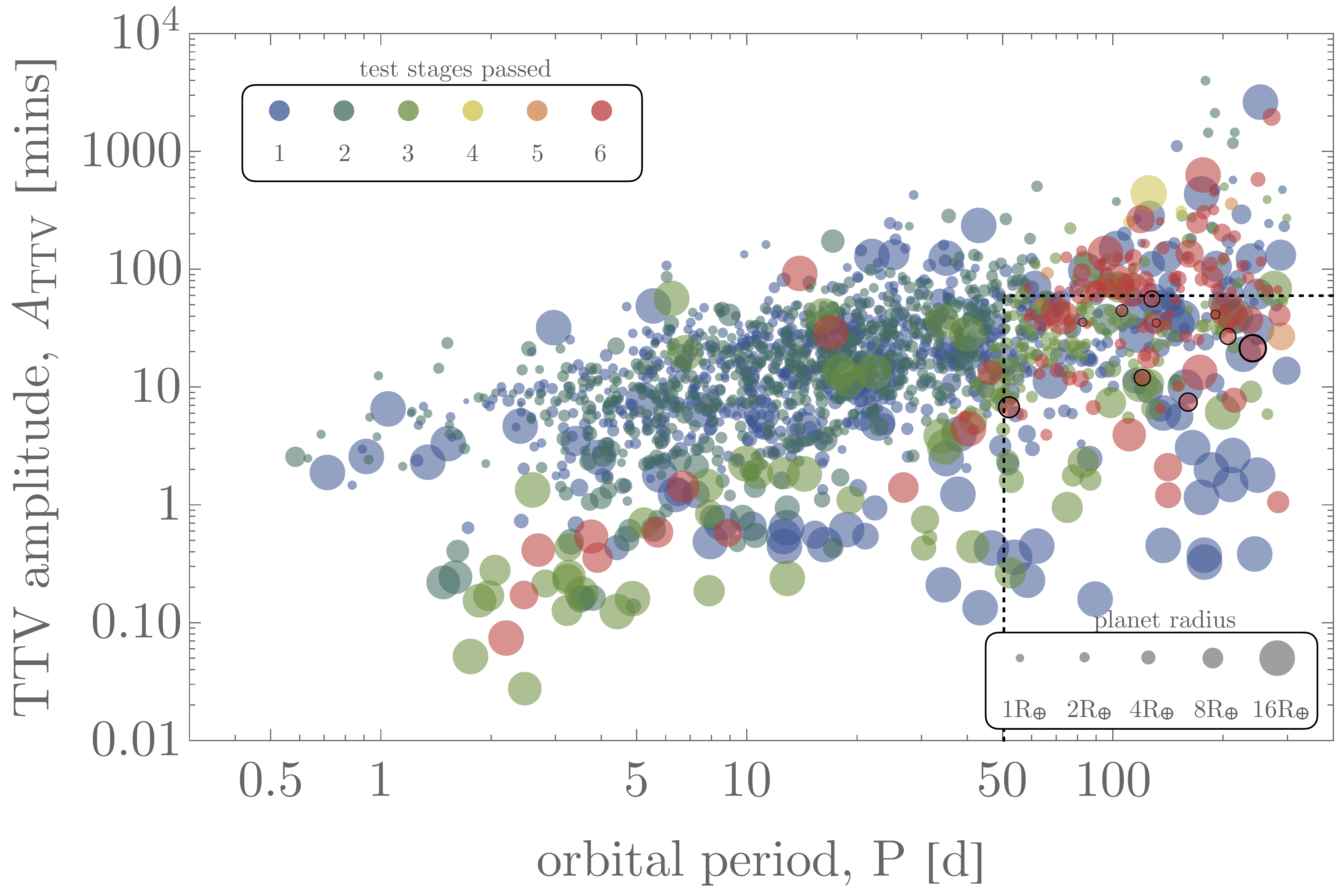}
\caption{
1822 KOIs for which we identify a $\Delta\mathrm{BIC}>0$ periodic
TTV signal of less than 20\,cycles periodicity using the H+16
transit timing catalog (with outlier rejection preconditioning applied).
The different colored point show the survivors from various test stages
applied in Section~\ref{sec:filtering}, where the population drops
from $1822\to917\to314\to129\to126\to122$ (blue to red) in applying the
various tests. The final 22 used in Section~\ref{sec:indep} are highlighted
with black circles.
}
\label{fig:pop_plot}
\end{center}
\end{figure*}

On 26th December 2021, we downloaded the NEA and flagged any KOIs for which
either the ``Exoplanet Archive Disposition'' \textit{or} the ``Disposition
Using Kepler Data'' columns listed ``FALSE POSITIVE''.  All such KOIs were
removed from our sample. At the same time, we also remove any KOIs residing
a multi-system. The justification for this is simply that single-planet
systems exhibiting TTVs demand additional mass in the system to explain
the data, whereas multi-planet data does not necessarily. For this reason,
we consider single-planet TTVs more interesting (for our purposes) than
multis. Applying these two criteria reduce the sample from 1822 KOIs to
917. Figure~\ref{fig:pop_plot} shows the TTV amplitudes versus orbital
periods of the sample before and after applying this cut, to visualise
how the population is sculpted by this filter (as well as subsequent
filters in this section).

\subsection{Remove TTVs above the exomoon ceiling}
\label{sub:ceilingcut}

For our third filter, we exploit the ``impossible moons'' argument of
\citet{impossible:2020}. In that work, it is shown that exomoons that are
bound to be i) within the Hill sphere of the their parent planet, and ii)
less massive than their parent planet, will have a well-defined maximum
possible TTV amplitude. Thus, any TTV amplitudes observed above this ceiling must
be non-exomoon in nature.

To calculate the ceiling, we first take the NEA reported maximum
\textit{a-posteriori} (MAP) ratio-of-radii (primarily sourced from the DR25
analysis of \citet{thompson:2018}) and multiply by it by the MAP stellar
radius reported in the homogeneous stellar catalog of \citet{berger:2020}.
These planetary radii are converted to a low- and high-estimate of the
planetary mass using a modified version of \forecaster\ \citep{chen:2017}.
Specifically, since here we only need a point-estimate, we take the
deterministic form of \forecaster\ to express $R_P(M_P)$ up to the
Neptunian-Jovian transition point of 131.6\,$M_{\oplus}$. Since the
equation is monotonic up to here, one can simply invert $M_P(R_P)$. Beyond
this transition, the relation become degenerate and so we make two extreme
assumptions about the planetary mass - it is either fixed to
131.6\,$M_{\oplus}$ (a Saturnian low-ball estimate) or to the mass
corresponding to the Jovian-stellar transition at 0.08\,$M_{\odot}$
(a brown dwarf high-ball estimate). With this, we are able to convert
the reported radii to masses, giving two extrema values for the
Jovian-sized worlds.

Equipped with a planetary mass, we can now estimate the maximum stable
exomoon mass using Equation~(9) of \citet{barnes:2002} - which considers
the tidal migration of a moon around a planet. This is technically done
twice, for the low- and high- mass estimates (although these estimates
are identical for sub-Jovian planets), and where all other
parameters are kept consistent from before (e.g. the \citet{berger:2020}
stellar parameters are used). This expression also requires
an estimate for the tidal value $(k_{2p}/Q_P)$, which is interpolated
based off the Solar System empirical $R$-$(k_{2p}/Q_P)$ relation following
\citet{teachey:2017} (see their Figure~1). We then compute the corresponding
TTV amplitude of the system using the model of \citet{kipping:2009a}
and compare the value to the amplitude found from the peak of the
LS periodogram. If the observed amplitude exceeds the ceiling, as
derived using \textit{either} of the low-/high-ball mass estimates,
then the KOI is rejected as a plausible exomoon signal. This filter
reduces our sample down from 917 KOIs to 314.

\subsection{Robustness test of the statistical preference}

In test 1, each signal was filtered to have a statistical preference for a
periodic TTV rather than a linear ephemeris. That test was conducted using the
BIC metric, but in isolation that metric can be misleading; in particular
because of the possible presence of outliers or additional stochastic noise not
captured by the formal TTV uncertainties. Although some outlier rejection was
applied during the LS periodogram (see Section~\ref{sub:LSperiodogram}), it
is possible outliers several times greater than the TTV scatter could persist.

To address this, we here employ a fourth filter, which ensures the statistical
preference is robust against these possibilities. To check against a single
dominant outlier, we iteratively dropped out a TTV point from the time series
and repeat the sinusoidal fit at the putative period. Cycling through all
possible permutations, we demand that the BIC on these truncated time series
always favours the periodic model. In this way, our claim for statistical
preference is robust against any one single point being an outlier.

A possible weakness of this test is that it truly only tests for robustness
against a single outlier, but multiple outliers could also deceive us into
claiming a spurious signal. Accordingly, we also demand that the $\Xi^2$ of the
sinusoidal fit residuals are lower than that of the linear ephemeris. As noted
earlier (see Section~\ref{sub:LSperiodogram}), this is a median-based version
of the $\chi^2$ and thus remains robust up to a ${<}50$\% contamination fraction
from outliers.

Another possibility is that there are no outlier per se, but that the timing
uncertainties are globally underestimated. As an example, starspots can induce
spurious TTVs \citep{ioannidis:2016} that would not be captured by the formal
uncertainties of H+16, since they did not employ a spot model in their
inference. As a test for this, let us assume that the periodic model is correct
but that the errors are too small. In that case, we would expect the $\chi^2$
of the periodic model with correctly adjusted uncertainties to approximately
equal the number of degrees of freedom. We thus scale the errors to satisfy
this condition\footnote{We also impose that this scaling factor cannot be
less than unity (i.e. we never shrink the formal uncertainties).} and then
recompute the $\Delta\mathrm{BIC}$ between the two competing models. Robust
models should still have a preference for the periodic model even after this
rescaling.

\begin{figure*}
\begin{center}
\includegraphics[width=17.4cm,angle=0,clip=true]{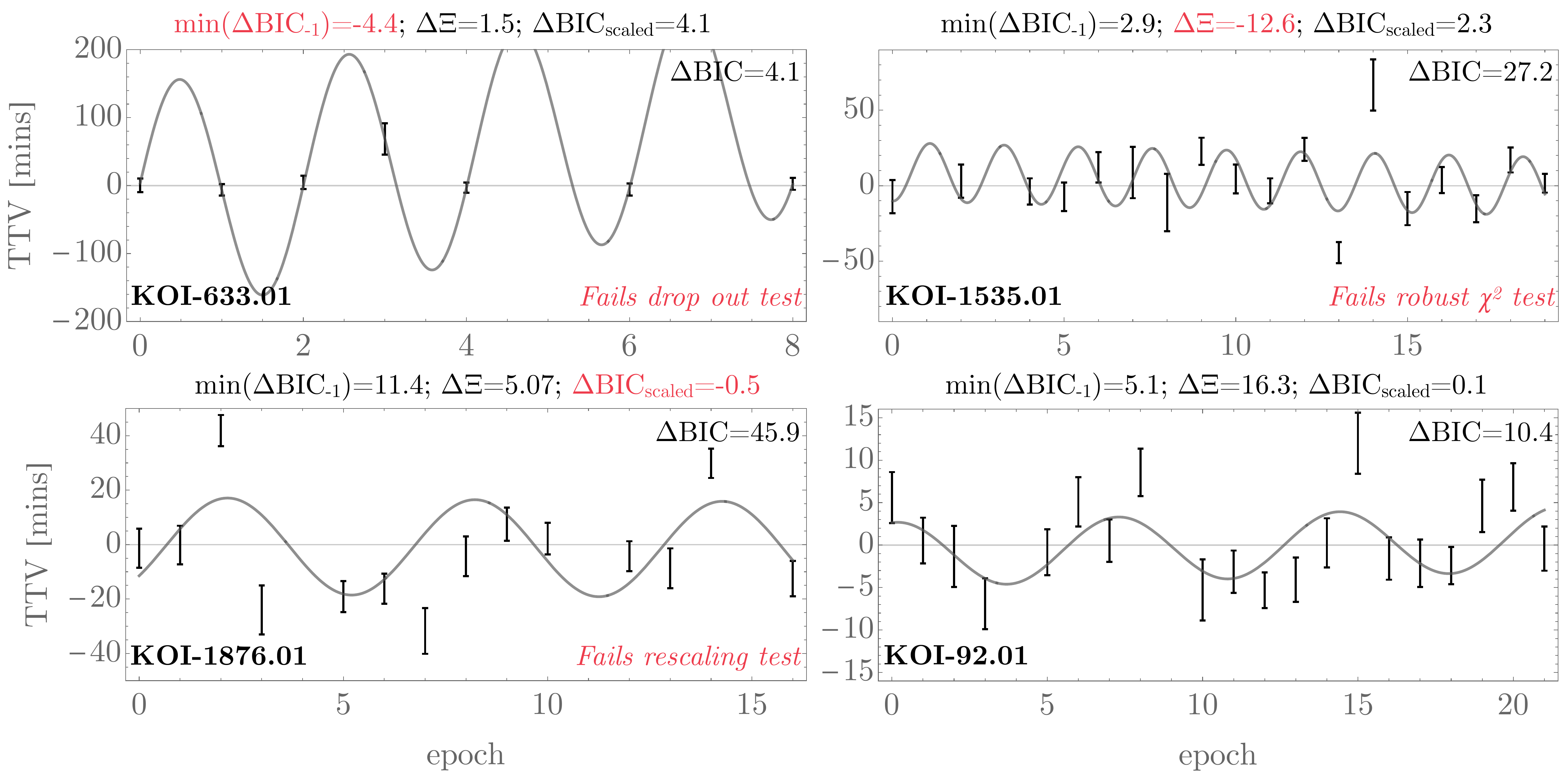}
\caption{
Top-left, top-right and lower-left panels show three examples of TTVs
derived from the H+16 the transit times that fail our significance
robustness tests. Each case highlights in red the specific test that
was failed. The lower-right panel shows an example of a KOI that passes
all tests. Each panel includes the best-fitting sinusoidal signal as
derived from a LS periodogram search.
}
\label{fig:rejected_examples}
\end{center}
\end{figure*}

Of the 314 remaining KOIs, 186 passed the drop-out test, 239 pass the $\Xi^2$
test and 243 pass the rescaling test. Applying all three together reduces our
sample of 314 KOIs to 129. To illustrate these tests in action,
Figure~\ref{fig:rejected_examples} shows three cases where just one of these
tests were failed (KOIs-633.01, 1535.01, 1876.01), and a fourth case where all
tests were passed (KOI-92.01).

\subsection{Test for suspicious phase clustering}

For a given KOI, one can fold the TTVs upon the peak period derived from
LS periodogram and inspect the distribution of the resulting phase-folded
TTVs. In general, there is no reason why a real signal would feature
clustering in phase-space, but this could be consistent with a spurious
signal. This can occur if the candidate period lies close to an aliasing
of the window function, leading to a chopping effect in the sampling.
Such suspicious cases can be flagged by simply phase-folding the TTVs
upon the best period and employing the classic Kolmogorov-Smirnov (KS) test
\citep{kolmogorov:1933,smirnov:1948}. We evaluate the $p$-value from the
KS test against a uniform distribution to flag KOIs with surprising
phase distributions. This test results in just three suspicious cases
($p<0.05$), including KOI-5942.01 which we highlight as an example in
Figure~\ref{fig:bad_phases}. With these three KOIs dropped, we are left
with 126 KOIs.

\subsection{Test for adequate phase coverage}

Along the same lines, the phase clustering could be quasi-uniform but have poor
coverage in the phase-folded space. To quantify this, we define a ``coverage
metric'', $C$, which is equal to the difference between the maximum and minimum
phase-folded TTV phase. Small $C$ values are clearly suspicious, implying
the observed TTVs only span a relatively small fraction of the supposed signal.

To identify surprisingly low $C$ values more rigorously, for each KOI we
took the TTVs and folded them upon a random frequency uniformly distributed 
between the minimum and maximum periods used in our LS periodogram search.
After folding, we measure the $C$ metric and then repeat until 1000 such
samples have been generated. We then evaluate the $p$-value of the real $C$
score by comparing to this distribution.

\begin{figure}
\begin{center}
\includegraphics[width=8.0cm,angle=0,clip=true]{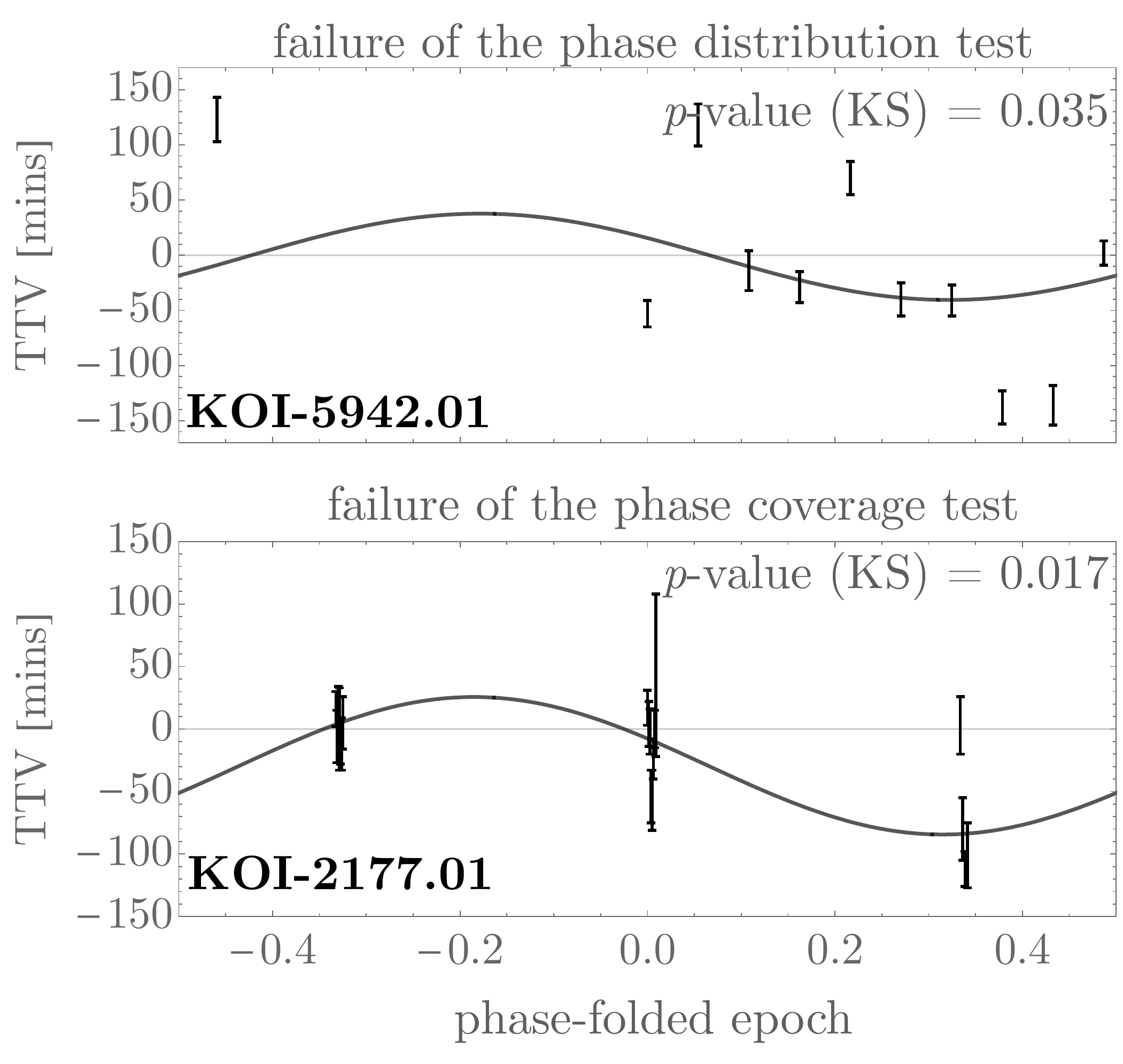}
\caption{
Top: an example of a TTV signal that fails our test for uniform phase coverage.
After phase-folding upon the candidate TTV period, the TTVs are disproportionally
occuring in the second-half of the diagram. Lower: an example of a TTV signal
that fails our phase coverage test. Given the generous number of points, it's
highly unlikely the phase-folded TTVs would only span ${\sim}70$\% of the
available window, and visual inspection confirms the spurious nature of
this signal. In both cases, the blue line shows the fitted candidate TTV signal.
}
\label{fig:bad_phases}
\end{center}
\end{figure}

Four KOIs were found to have suspiciously low $C$ scores ($p<0.05$), with the
example of KOI-2177.01 shown in Figure~\ref{fig:bad_phases}. KOI-2177.01
is particularly interesting to highlight because inspection of the phase
folded TTVs reveals a clearly spurious signal, but one that unexpectedly passed
the previous phase distribution test. Indeed, with the phase distribution test,
the $p$-value of the KS test against a uniform distribution was $0.081$,
but with the coverage test it finally gets flagged with $p=0.017$. Such cases
highlight that no battery of tests is perfect, but in unison a large number
of spurious cases can be trimmed, saving timing in the later analyses. With
these four KOIs removed, we are left 122 KOIs.

\subsection{Final downsampling}

For our final downsampling, we more aggressively hone in on potentially
significant, highly-exomoon like signals. To this end, we apply a much
stronger statistical cut that $\Delta(\mathrm{BIC})>10$ and only accept
exomoon corridor KOIs such that $\Pttv<4$\,cycles (corresponding to the 50th
percentile of expected exomoon periods; \citealt{corridor:2021}). We also
eliminate any signals with TTV amplitudes in excess of one hour, $\leq4$
available epochs and planets with periods less than 50\,days. We find that
11 of the 122 KOIs fall within this range and define our H+16 exomoon-like
TTV candidates.

We note that 120 of the 122 satisfy our revised statistical threshold, which is
perhaps not surprising given the battery of robustness tests such signals have
thus far survived. However, we caution against treating these as ``detections''
in the absence of an independent light curve analysis. For as the next section
will show, most of our 11 exomoon corridor candidates do not survive an
independent analysis. This final subset is highlighted with black circles in
Figure~\ref{fig:pop_plot}, and are listed in Table~\ref{tab:KOIs} (although
KOI-5825.01 is excluded since our later fits are unable to constrain its
transit times, see Section~\ref{sub:comparison}).

\section{Independent Analyses of the Survivors}
\label{sec:indep}

Having identified KOIs that appear to have significant TTV signals within
the exomoon corridor based upon the H+16 catalog, we next challenge the
veracity of that claim through an independent light curve analysis.

\subsection{Light curve detrending}
\label{sub:detrending}

We begin by performing an independent light curve detrending of the \kepler\
time series photometry; specifically here we use the long-cadence (LC) data.
The light curve treatment is almost identical to that of \citet{k1708b:2022}
and so we direct the reader there for details. In brief, the light curves are
detrended on an epoch-by-epoch basis using four distinct algorithms (\cofiam,
\polyam, \local\ and \gp) applied to two distinct data products (SAP and PDC).
For each transit of each KOI, this produces eight light curves which are then
tested to see if they behave consistently with Gaussian noise, and if so are
then combined together through ``method marginalisation''. This final stage
takes the median of the $\leq8$ photometric points at each time stamp, and
propagates a robust standard deviation estimate between them into the formal
uncertainties. The final method marginalised light curves are used for fitting.

The only difference between our treatment here and that of \citet{k1708b:2022}
is that the H+16 transit times are used (where available) to define the
mid-point of each epoch. This is helpful since all algorithms need to mask the
transit signal, which of course depends upon the temporal location of the
transit.

\subsection{Light curve fits}
\label{sub:fits}

For each KOI, two models are regressed to the detrended data. Model
$\mathcal{P}$ assumes a linear ephemeris for the planet whereas model
$\mathcal{T}$ assumes each epoch has a unique time of transit minimum,
$\tau_e$. Both models use the \citet{mandel:2002} algorithm, combined with the
\citet{danby:1988} solver for the Kepler equation, numerical resampling of the
finite integration time following the method of \citet{binning:2010} and the
re-parameterised quadratic limb darkening coefficients of \citet{q1q2:2013}.

The regression was performed using the multimodal nested sampling code
\multi\ \citep{feroz:2009} using 4000 live points. Our fits adopted a uniform
prior for the ratio-of-radii, $p$, between 0 and 1, a uniform prior for the
impact parameter, $b$, between 0 and 1, a log-uniform prior for stellar
density, $\rho_{\star}$, between $10^{-3}$\,g\,cm$^{3}$ to
$10^{+3}$\,g\,cm$^{3}$, a uniform prior for period, $P$, and time of transit
minimum, $\tau$, between $\pm1$\,day of that quoted by the NEA, and uniform
priors for limb darkening coefficients $q_1$ and $q_2$ between 0 and 1. In
total then, model $\mathcal{P}$ had 7 free parameters. For model $\mathcal{T}$,
the parameters are the same except the period is not fitted and simply fixed
to the NEA MAP value. Similarly, there is no reference epoch, but instead a
unique transit time for each epoch, $\tau_i$. This gives $5+N$ parameters for
model $\mathcal{T}$, where $N$ is the number of available epochs.

\multi\ substantially slows down beyond 20-25 free parameters, and thus
if $N\gg1$, this can lead to excessive computational times. KOIs-1355.01 and
KOI-2992.01 have 27 and 16 available epochs respectively, sufficient that we
chose to split the $\mathcal{T}$ model up into 2 and 3 segments respectively,
following \citet{teachey:2018}. The downside of this is that the global shape
parameters, such as $b$ and $\rho_{\star}$, are not self-consistent between
the segments and not as precise as they could be, since they are conditioned
upon only a fraction of the data. However, since our primary interest is the
transit times themselves, which have little covariance with the other terms
\citep{carter:2008}, this trade-off was considered acceptable.

\subsection{TTV analysis}
\label{sub:ttvanalysis}

The posteriors obtained in Section~\ref{sub:fits} may be used to define
credible intervals for the transit times. We do so using the median
and 1\,$\sigma$ quantiles.

Equipped with our independently derived transit times, we now proceed to
evaluate if the candidate exomoon corridor signals persist in this cross
check. The first step is to remove any possible outlier transit times.

Since have fit the data with a second model that explicitly assumes
a linear ephemeris, we take the MAP ephemeris from model $\mathcal{P}$
of each planet as a preliminary reference ephemeris to define preliminary
TTVs. From these, we calculate the standard deviation as a measure of the
expected level of dispersion, against which we could look for outliers.
However, even the standard deviation is itself a noisy estimate and so we add
5 standard deviations of the standard deviation onto this estimate
to obtain a conservative maximum expected TTV deviance. Epochs where both
components of the asymmetric uncertainty exceed this threshold were
removed, since have weak constraining power and are likely driven by a
partial/anomalous transit profile. We repeated this calculation on the
H+16 transit times to produce a later fair comparison.

We next turn to repeating the periodogram exercise. Since our own transit
times have asymmetric uncertainties, we modify our weighted linear least 
squares scheme to accommodate such asymmetries. The introduction of a
Heaviside Theta function necessary to do this leads to a now non-linear
problem, and thus the best-fit is obtained using a Nelder-Mead minimisation
\citep{nelder:1965} of the $\chi^2$ function. The periodogram uses a
period grid defined in the same way as in Section~\ref{sub:LSperiodogram},
and similarly two models are regressed to the raw transit times (not the TTVs);
a 2-parameter linear ephemeris model and a 4-parameter\footnote{Although
four parameters are formally in the regression, recall that this regression
sweeps across a grid of TTV periods too and we extract the maximum peak,
hence defining an additional dimension of freedom.} linear ephemeris
plus sinusoidal variation model. 

\subsection{Comparison to H+16 TTVs}
\label{sub:comparison}

\begin{figure*}
\begin{center}
\includegraphics[width=17.4cm,angle=0,clip=true]{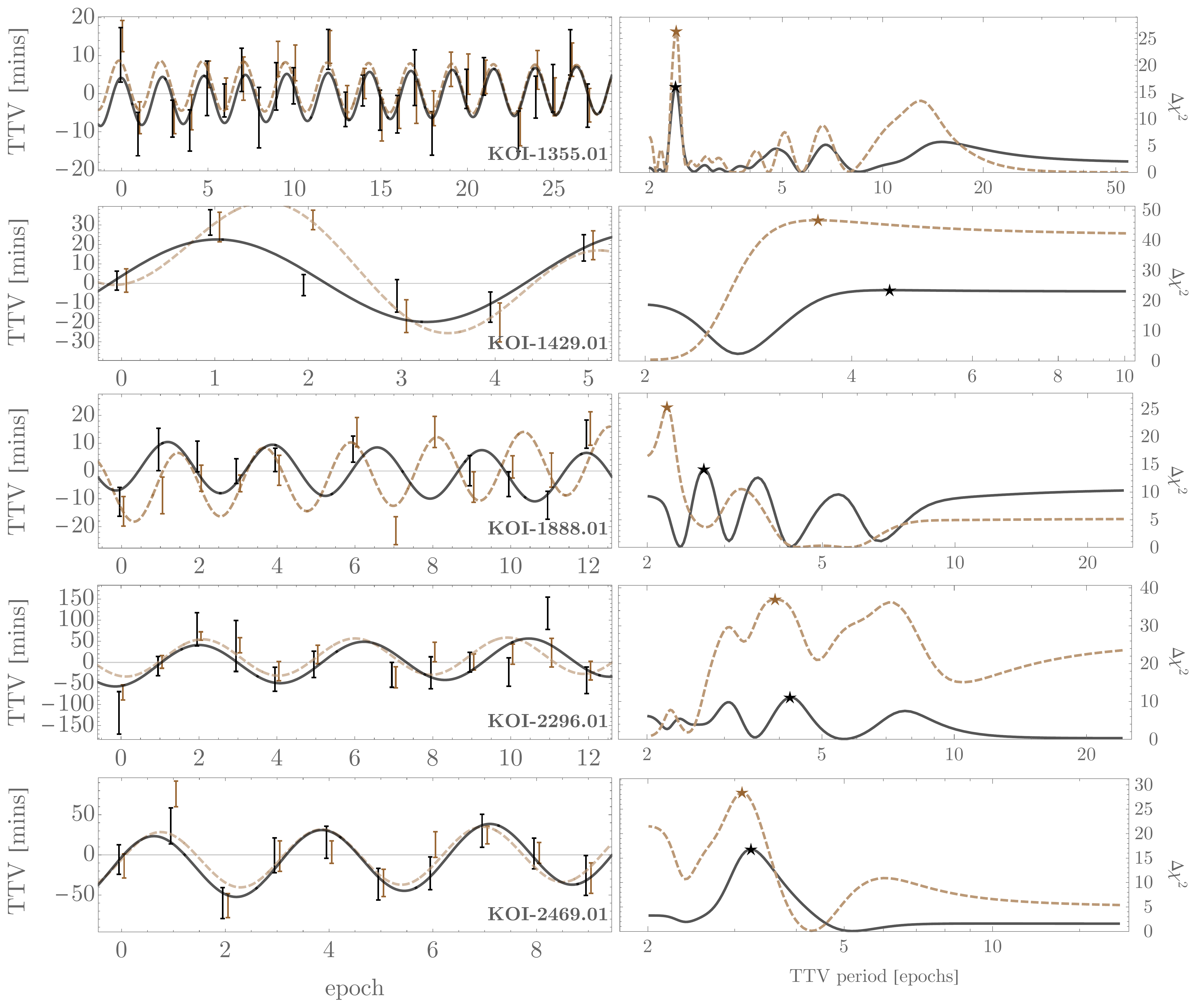}
\caption{
TTVs (left-panel) for 5 KOIs (labeled in the corners) ostensibly in the exomoon
corridor to high significance from the H+16 catalog. Black points are those
derived in this work from a full independent analysis, brown are those derived
by H+16. With the same colours we overlay the best fitting sinusoid for each.
Right panel shows the periodogram for each, where the best-fit (maximum
likelihood) solution is highlighted with a star.
}
\label{fig:gridA}
\end{center}
\end{figure*}

\begin{figure*}
\begin{center}
\includegraphics[width=17.4cm,angle=0,clip=true]{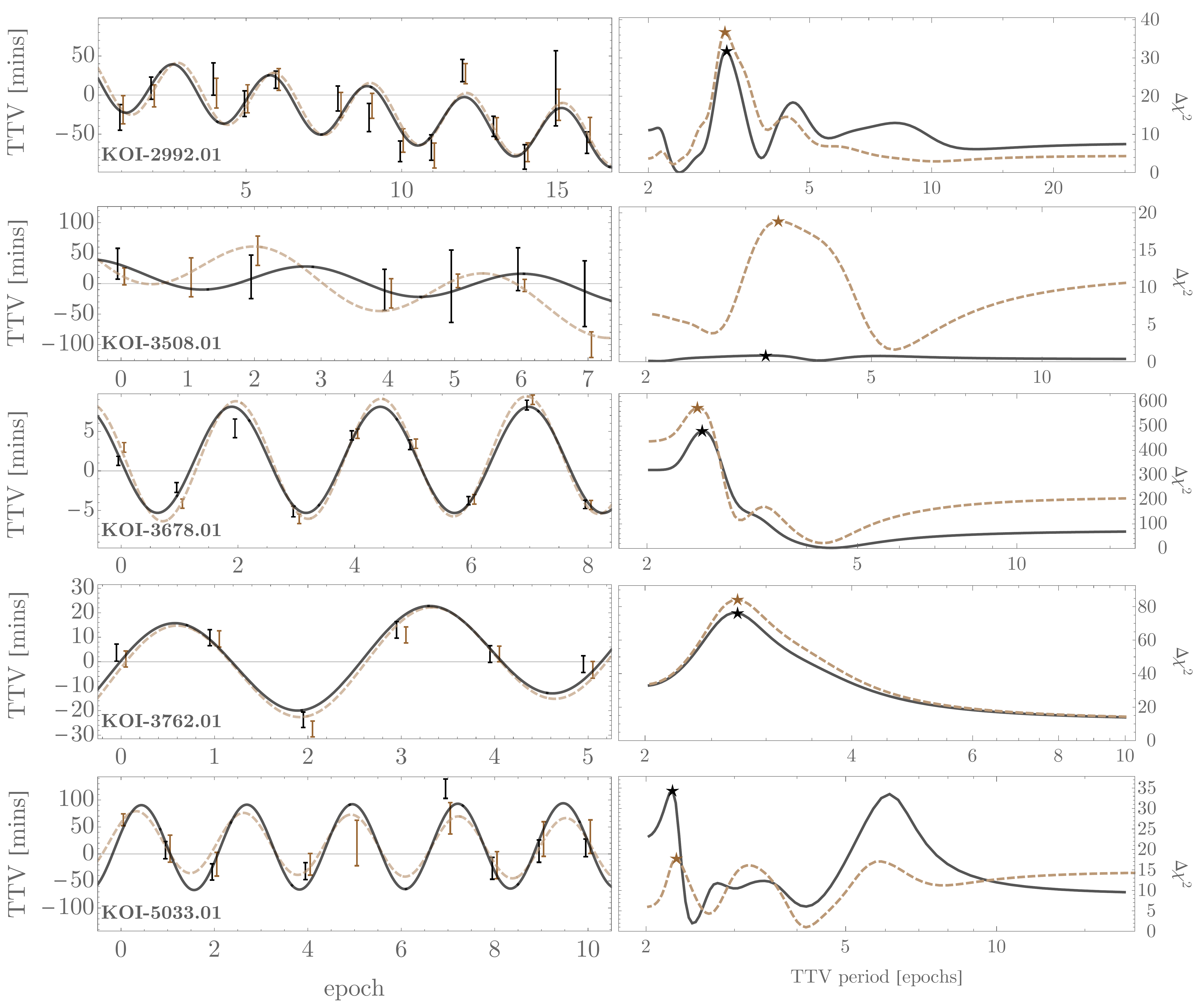}
\caption{
Same as Figure~\ref{fig:gridA} for the other 5 KOIs.
}
\label{fig:gridB}
\end{center}
\end{figure*}

Figures~\ref{fig:gridA} \& \ref{fig:gridB} present the TTVs of 10 of the 11
remaining KOIs (left panels). KOI-5825.01 was removed here since our own fits
were unable to constrain the transit times at all, and every fit posterior
returned the prior. The left-hand side figure panels also show the H+16
transit times for comparison in brown and slightly (artificially) offset to the
right for clarity. Overlaid is the best-fitting sinusoidal signal from the
periodogram search, for both data inputs.

This comparison reveals that in most cases there is good agreement concerning
the locations of the best-fitting transit times, although exceptions certainly
are present such as epoch 2 of KOI-1429.01, or epoch 11 of KOI-2296.01.
Additionally, H+16 occasionally includes epochs that our own approach rejected,
generally because the light curve quality was unacceptably poor; for example
epochs 7 and 8 of KOI-1888.01 are present with H+16 but absent here. In that
case, one can see that these epochs have a large impact on the retrieved
sinusoid, leading to a much shorter period signal approaching the Nyquist rate.

Across the ensemble, the uncertainties are often quite distinct between our own
measurements than that of H+16. Cross-matching epochs for which both data sets
provide a transit time and ratioing their mean uncertainties, we find that our
uncertainties are larger than that of H+16 in 65.6\% of cases, with a median
ratio of 1.09x and a mean ratio of 1.27x. The fact that H+16 has a tendency for
smaller errors has an important consequence - the periodograms typically return
higher significances than that derived using our own times. This is
evident from the right panels of Figures~\ref{fig:gridA} \& \ref{fig:gridB},
where the H+16 periodogram peak exceeds that of our own in 9 out of 10 cases.
As a result, although all 10 have $\Delta\mathrm{BIC}>10$ using H+16 transit
times, that number halves with the times derived here (KOIs-1429.01, 2992.01,
3678.01, 3762.01 \& 5033.01). Some possible explanations for the discrepancies
between H+16's transit times and that derived here are offered in
Section~\ref{sec:discussion}.

\subsection{Scrutinizing the candidacy of the remaining signals}

The work presented here seeks to identify exomoon-like signatures using TTVs
alone. Such an effort was previously made by \citet{fox:2021}, who claimed
the existence of numerous new exomoon candidates by inspection of the
H+16 TTV catalog. The claim was challenged in two independent papers, one
which highlighted the implausible ability of many of these to retain large
exomoons due to tidal evolution \citep{quarles:2020} and another which
showed that the KOIs failed a trifecta of tests designed to test to assess
their candidacy \citep{sixmoons:2020}. The latter paper also highlighted
the importance of independently analysing the \kepler\ light curves, rather
than adopting the H+16 catalog values as final. The three tests proposed
serve as a benchmark standard in the current literature and thus we elected
to adopt them here to scrutinise our remaining putative KOI signals. Afterall,
if TTV exomoon candidates were killed using these tests in
\citet{sixmoons:2020}, then any compelling signals claimed here should at
least pass those same tests, which centre around three questions:

\begin{itemize}
\item[{\textbf{Q1]}}] Are there statistically significant TTVs?
\item[{\textbf{Q2]}}] Is there a statistically significant periodic TTV?
\item[{\textbf{Q3]}}] Do the observations support a statistically significant non-zero moon mass?
\end{itemize}

\subsection{Q1 - Significant TTVs?}

At this point, the reader might suspect Q1 has already been convincingly
answered. Afterall, the five KOIs remaining have been subject to a battery of
significance tests already, including a complete re-analysis of the light
curve data. However, all of these tests exploit the derived transit times,
which are really a meta-product, rather than the photometry itself. Further,
they frame the BIC in terms of raw degrees of freedom, rather the physical
degrees of freedom. To address this, Q1 takes the maximum likelihood from
the photometric light curve fits of models $\mathcal{P}$ and $\mathcal{T}$
and compares those, rather than metrics based on the TTV meta-products.
Following \citet{sixmoons:2020} (see Section 3.1.4), the number of degrees of
freedom is based upon the physical models that are needed to explain the
data rather than that of the more ad-hoc case of simply assigning every
epoch a new degree of freedom.

We applied this test to our 10 remaining KOIs, despite the fact 5 of them have
already been shown to be spurious. We chose to do all 10 to provide some
greater contextual examples of these tests in action. The results are shown in
Table~\ref{tab:KOIs}, where one can see that only 3 KOIs pass this simple test,
KOIs-3678.01, 3762.01 and 5033.01. We note that all three of these are members
of the five KOIs that survived our independent light curve analysis test in
Section~\ref{sub:comparison}. KOI-3678.01 particularly stands out with
an enormous significance score of $\Delta\mathrm{BIC}=207.9$, which is
particularly remarkable because the TTV amplitude is only 6.7\,minutes
(see Table~\ref{tab:KOIs}).

\subsection{Q2 - Significant periodic TTV?}

In Q2, we follow the cross-validation tests of \citet{sixmoons:2020} using the
same 20\% hold-out set (rounded to an integer number of epochs). Since
$0.2 \times n < 1.5 $ for all $n \leq 7$, then KOIs with 7 or fewer useful
epochs would only have a single epoch held-out, occurring for KOIs-1429.01,
3762.01 and 5033.01. In this test, the entire periodogram process is repeated
a large number of times, which itself scans over a large number of frequencies.
To maximise the efficiency of our cross-validation search then, the number of
iterations of hold-out sets was set to $\binom{n}{k}$, where $n$ is the number of
epochs available and $k$ is the size of the holdout set. In this way, we cycle
through all possible hold-out sets once. The only exception to this was
KOI-1355.01, for which $\binom{n}{k} = \binom{24}{5} = 42504$ which was simply
too large to run in the time available. Instead, here, we simply did $425$ as a
1\% representative sample.

The cross-validation test essentially demands that the best-fitting model has
useful predictive power, a basic expectation of a genuine physical model. Our
criterion here is that $>50$\% of the predictions using the sinusoidal TTV
model must be superior to that of the linear ephemeris model, in a $\chi^2$
sense. We are careful here to keep track of the asymmetric uncertainties in the
transit times too in defining our $\chi^2$ metric. The result is that four KOIs
pass this test, KOI-1355.01, 2469.01, 3508.01 and 3678.01. At this point, only
KOI-3678.01 now has a clean sweep of successful vetting tests.

\subsection{Q3 - Significant non-zero moon mass?}

As our third and final test, we fit (using \multi) a planet+moon photodynamical
light curve model using \luna\ \citep{luna:2011} to the light curves. In this
fit, we turn off the usual likelihood penalty function in our code that rejects
any samples with unphysical satellite densities. This allows the fit to explore
very low satellite masses without penalty and thus the moon mass posterior is
able to peak at zero in cases where no evidence for a moon-like dynamical
signature is found. We then assess the posterior density at zero with a
Savage-Dickey ratio \citet{dickey:1971}, to evaluate the Bayes factor of a
non-zero mass exomoon model against the null (zero mass signal), denoted as
$K_{\mathcal{M}:\mathcal{X}}$. We consider a KOI to pass this test if
$K_{\mathcal{M}:\mathcal{X}} > 10$, denoting ``strong evidence'' on the
\citet{kass:1995} scale.

The major benefit of this test over simple sinusoidal signal check is that
exomoon TTVs are not always strict sinusoids \citep{rodenbeck:2020}. Moon
acceleration during the transit can distort the shape of the signal and
thus invalidates the assumptions of the analytic TTV theory (which predicts
a sinusoid). Two KOIs pass this final test, but of most relevancy is that
KOI-3678.01 completes a clean sweep of successful tests with this final
challenge. At this point, we are confident that KOI-3678.01 is a genuine
TTV signature with a periodicity, amplitude and shape consistent with that of
an exomoon. We discuss this object in more detail in the next section.

It is also worth remarking that amongst the other KOIs, KOI-3762.01 comes
very close to a full sweep as well, failing only the cross-validation test.
Looking at the TTVs shown in Figure~\ref{fig:gridB}, one can see that
the signal is strongly dependent upon a single epoch, epoch \#2 in the
figure. That one point is down, as found by both our work and H+16. The
importance of that single point, and the paucity of epochs here (just 6)
makes this a challenging KOI to work with. On this basis, we encourage
future transit timing observations of KOI-3762.01, as it's unclear how
robust our cross-validation results are in such a challenging case.

\newpage
\newgeometry{left=1cm,top=7cm} % modify this if you need even more space
\begin{landscape}
\thispagestyle{empty}
\begin{table*}
\caption{
List of the various test metrics applied to the 11 KOIs which
appear to exhibit an exomoon corridor signal using the H+16
catalog. Columns 2-4 list a pair of numbers, the first as
derived using H+16 transit time and the second using our own
analysis. Columns 5-8 list the trifecta of higher-level TTV
tests previously proposed by \citet{sixmoons:2020}.
} % title of Table
\centering % used for centering table
\begin{tabular}{c c r r r c r r c} % centered columns (8 columns)
\hline\hline %inserts double horizontal lines
KOI & \vline &
$A_{\mathrm{TTV}}$\,(mins) &
$P_{\mathrm{TTV}}$\,(epochs) &
$\Delta\mathrm{BIC}$ & \vline &
$\Delta\mathrm{BIC}_{\mathrm{physical}}$ \textbf{[Q1]} &
\% of good TTV predictions \textbf{[Q2]} &
$\log K_{\mathcal{M}:\mathcal{X}}$ \textbf{[Q3]} \\ [0.5ex] % inserts table
%heading
\hline % inserts single horizontal line
1355.01	& \vline &
$\{6.51,6.26\}$ [\checkmark,\checkmark] & % Attv
$\{2.41,2.39\}$ [\checkmark,\checkmark] & % Pttv
$\{16.8,6.4\}$ [\checkmark,\text{\sffamily X}] & \vline & % BIC
-11.3 [\text{\sffamily X}] &
(287/425$^{\dagger}$) 67.5\% [\checkmark] & -1.3 [\text{\sffamily X}] \\
1429.01	& \vline &
$\{27.13,21.98\}$ [\checkmark,\checkmark] & % Attv
$\{3.57,4.55\}$ [\checkmark,\text{\sffamily X}] & % Pttv
$\{41.3,18.1\}$ [\checkmark,\checkmark] & \vline & % BIC
-8.3 [\text{\sffamily X}] &
(0/6) 0.0\% [\text{\sffamily X}] & 0.1 [\text{\sffamily X}] \\
1888.01	& \vline &
$\{11.87,8.96\}$ [\checkmark,\checkmark] & % Attv
$\{2.22,2.69\}$ [\checkmark,\checkmark] & % Pttv
$\{25.3,7.2\}$ [\checkmark,\text{\sffamily X}] & \vline & % BIC
-11.8 [\text{\sffamily X}] &
(5/45) 11.1\% [\text{\sffamily X}] & 0.7 [\text{\sffamily X}] \\
2296.01	& \vline &
$\{43.60,47.20\}$ [\checkmark,\checkmark] & % Attv
$\{3.91,4.23\}$ [\checkmark,\text{\sffamily X}] & % Pttv
$\{29.5,3.6\}$ [\checkmark,\text{\sffamily X}] & \vline & % BIC
-14.2 [\text{\sffamily X}] &
(0/66) 0.0\% [\text{\sffamily X}] & -1.3 [\text{\sffamily X}] \\
2469.01	& \vline &
$\{35.25,39.75\}$ [\checkmark,\checkmark] & % Attv
$\{3.10,3.24\}$ [\checkmark,\checkmark] & % Pttv
$\{21.5,9.9\}$ [\checkmark,\text{\sffamily X}] & \vline & % BIC
-14.1 [\text{\sffamily X}] &
(24/45) 53.3\% [\checkmark] & -1.1 [\text{\sffamily X}] \\
2992.01	& \vline &
$\{36.40,34.42\}$ [\checkmark,\checkmark] & % Attv
$\{3.09,3.13\}$ [\checkmark,\checkmark] & % Pttv
$\{28.9,23.9\}$ [\checkmark,\checkmark] & \vline & % BIC
+1.3 [\text{\sffamily X}] &
(89/364) 24.4\% [\text{\sffamily X}] & -0.2 [\text{\sffamily X}] \\
3508.01	& \vline &
$\{41.36,21.92\}$ [\checkmark,\checkmark] & % Attv
$\{3.43,3.26\}$ [\checkmark,\checkmark] & % Pttv
$\{13.1,-4.5\}$ [\checkmark,\text{\sffamily X}] & \vline & % BIC
-18.6 [\text{\sffamily X}] &
(5/6) 83.3\% [\checkmark] & -0.5 [\text{\sffamily X}] \\
3678.01	& \vline &
$\{7.50,6.70\}$ [\checkmark,\checkmark] & % Attv
$\{2.49,2.54\}$ [\checkmark,\checkmark] & % Pttv
$\{569.2,474.2\}$ [\checkmark,\checkmark] & \vline & % BIC
+207.9 [\checkmark] &
(35/36) 97.2\% [\checkmark] & 33.8 [\checkmark] \\
3762.01	& \vline &
$\{56.52,19.57\}$ [\checkmark,\checkmark] & % Attv
$\{2.73,2.73\}$ [\checkmark,\checkmark] & % Pttv
$\{78.9,70.9\}$ [\checkmark,\checkmark] & \vline & % BIC
+13.4 [\checkmark] &
(0/6) 0.0\% [\text{\sffamily X}] & 34.5 [\checkmark] \\
5033.01	& \vline &
$\{56.52,78.72\}$ [\checkmark,\text{\sffamily X}] & % Attv
$\{2.30,2.26\}$ [\checkmark,\checkmark] & % Pttv
$\{11.2,28.6\}$ [\checkmark,\checkmark] & \vline & % BIC
+25.7 [\checkmark] &
(1/7) 14.3\% [\text{\sffamily X}] & -1.92 [\text{\sffamily X}] \\ [1ex]
\hline\hline %inserts single line
\end{tabular}
\label{tab:KOIs} % is used to refer this table in the text
\end{table*}
\end{landscape}
\restoregeometry

%\newpage
%\begin{sidewaystable}
%\input{KOItab.tex}
%\end{sidewaystable}

\section{The Case of Kepler-1513b}
\label{sec:3678}

\subsection{System parameters}

KOI-3678.01 emerges as the only TTV candidate from our analysis for which
we can confidently determine it exhibits a significant exomoon-corridor TTV.
We note that this object was previously validated as Kepler-1513b
\citep{morton:2016} and we will switch to this monicker in what follows.
This does not mean no other \kepler\ candidates are in the corridor, since
numerous filters applied in this work may have excluded them (e.g. multis
were excluded and should also exhibit corridor signatures;
\citealt{teachey:2021}). An example of this would be Kepler-1625b, which only
has 3 \kepler\ epochs from which no clear TTVs are evident, but inclusion of a
fourth epoch from the Hubble Space Telescope reveals a significant TTV
\citep{teachey:2018} that appears consistent with the exomoon corridor despite
the significant degeneracies of only having four epochs in hand (see
\citealt{corridor:2021}).

Although one of our filters in Section~\ref{sec:filtering} screened for 
TTVs above the ceiling predicted due to an exomoon, leveraging the tidal theory
of \citet{barnes:2002}, it is worth re-visiting the question as to how
physically feasible it is for Kepler-1513b to have an exomoon. Afterall, it was
through this argument that \citet{quarles:2020} questioned the existence of
the purported exomoons claimed in \citet{fox:2021}, which also exclusively used
TTVs. In order to address this, we first need a set of fundamental parameters
for the star and planet, which is challenged by the fact there is no planetary
mass measurement in the previous literature.

To tackle this, we first used the \isochrones\ package \citep{morton:2015} to
derive posterior samples for the star. The stellar atmospheric parameters are
taken from P. Dalba (priv. comm.) using Keck HIRES and \SpecMatch\ \citep{petigura:2015,2017}, the parallax from \gaia\
DR3 \citep{gaia:2021}, and the \kepler\ apparent magnitude is used to constrain
the luminosity. This reveals the star to be a late G-type star, slightly
sub-Solar, with the parameters listed in Table~\ref{tab:3678}.

These posteriors samples are combined with those from the $\mathcal{T}$ light
curve fits (which recall account for TTVs) to produce physical parameters for
the planet itself, which are also listed in Table~\ref{tab:3678}. This reveals
Kepler-1513b to be around 75\% of a Jupiter-radius. Comparing to the empirical
mass-radius relation of \citet{chen:2017}, one can see that the planet is
likely a sub-Jovian mass object, but could also plausibly be a massive, highly
compressed brown dwarf. This is directly seen when we push the radius samples
through \forecaster\ \citep{chen:2017} giving a bimodal mass prediction.
RV observations using SOPHIE cannot detect Kepler-1513b's mass, but they do
constrain it to be ${<}1.43$\,$M_{\mathrm{Jup}}$ to 99\% confidence, thus
putting pressure on the second mode. This fact, combined with the lower
occurrence rate of more massive planets \citep{fulton:2021} and the preference
to not be overzealous in predicting large moons (which large planetary masses
inevitably allow for) led us to completely exclude the second mode which
delineates at $178$\,$M_{\oplus}$. Since each radius sample has a
corresponding and covariant forecasted mass sample, we are careful to apply
this down-filtering to the entire joint posterior samples to maintain the
correct covariance. This leaves us with a mass forecast of
$[14.8\,M_{\oplus},125.6\,M_{\oplus}]$ to 95.45\% confidence.

\subsection{Maximum stable exomoon mass}

It is now possible to more rigorously revisit the \citet{barnes:2002} tidal
limits for moon masses. To do so, we use our joint
posteriors including the truncated forecasted \citet{chen:2017} samples, the
$R$-$(k_{2p}/Q_P)$ relation of \citet{teachey:2017}, and plug them into to
Equation~(8) of \citet{barnes:2002}. The equation requires an assumed maximum
stable moon orbital radius, which we here take to be 0.4895 for a prograde moon
following \citet{domingos:2006}, and a lifetime, which we take to be 5\,Gyr.

From this we obtain a maximum stable moon mass of
$4.4_{-3.5}^{+13.4}$\,$M_{\oplus}$, with a 2\,$\sigma$ lower limit at
$0.32$\,$M_{\oplus}$. Accordingly, Kepler-1513b appears capable of supporting a
potentially terrestrial mass exomoon for 5\,Gyr.

\begin{table*}
\caption{
Credible intervals for various parameters of interest for Kepler-1513b
(KOI-3678.01).
} % title of Table
\centering % used for centering table
\begin{tabular}{c c c} % centered columns (8 columns)
\hline\hline %inserts double horizontal lines
Parameter & Definition & Credible Interval \\
\hline
& \textit{Observational stellar parameters} & \\
$K_P$ & \kepler\ apparent magnitude & $12.888\pm0.100$ \\
$T_{\mathrm{eff}}$ (K) & Stellar effective temperature & $5491\pm100$ \\
(M/H) (dex) & Stellar metallicity & $0.17\pm0.06$ \\
$\log g$ (dex) & Stellar surface gravity & $4.46\pm0.10$ \\
$\pi$ (mas) & Parallax & $2.8446\pm0.0134$ \\
\hline
& \textit{Fundamental stellar parameters} & \\
$M_{\star}$ ($M_{\odot}$) & Stellar mass & $0.943_{-0.037}^{+0.038}$ \\
$R_{\star}$ ($R_{\odot}$) & Stellar radius & $0.950_{-0.055}^{+0.078}$ \\
$\rho_{\star}$ (kg\,m$^{-3}$) & Stellar density & $1540_{-340}^{+350}$ \\
$L_{\star}$ ($L_{\odot}$) & Stellar luminosity & $0.74_{-0.10}^{+0.15}$ \\
\hline
& \textit{Transit fit parameters} & \\
$R_P/R_{\star}$ & Ratio-of-radii & $0.07882_{-0.00022}^{+0.00036}$ \\
$\rho_{\star,\mathrm{LC}}$ (kg\,m$^{-3}$) & Transit stellar density & $1304_{-53}^{+21}$ \\
$b$ & Impact parameter & $0.109_{-0.075}^{+0.095}$ \\
$q_1$ & Limb darkening parameter & $0.446_{-0.039}^{+0.040}$ \\
$q_2$ & Limb darkening parameter & $0.327_{-0.028}^{+0.031}$ \\
\hline
& \textit{TTV fit parameters} & \\
$P_P$ (days) & Orbital period & $160.884435_{-0.000064}^{+0.000062}$ \\
$\tau_0$ (BJD$_{\mathrm{UTC}}$-2,455,000) & Time of transit minimum & $110.50597_{-0.00031}^{+0.00032}$ \\
$\Pttv$ (cycles) & TTV periodicity & $2.556_{-0.018}^{+0.018}$ \\
$A_{\mathrm{TTV}}$ (mins) & TTV amplitude & $6.95_{-0.33}^{+0.32}$ \\
$\phi_{\mathrm{TTV}}$ (rads) & TTV phase & $3.212_{-0.074}^{+0.072}$ \\
\hline
& \textit{Estimated parameters} & \\
$M_{P,\mathrm{\forecaster}}$ ($M_{\oplus}$) & Forecasted (+truncated) planet mass & $48_{-21}^{+35}$ \\
$M_{S,\mathrm{max}}$ ($M_{\oplus}$) & Maximum stable moon mass & $4.4_{-3.5}^{+13.4}$ \\
$M_{S,\mathrm{TTV},\mathrm{min}}$ ($M_{\leftmoon}$) & Minimum implied moon mass from TTVs & $0.76_{-0.25}^{+0.33}$ \\ [1ex]
\hline\hline %inserts single line
\end{tabular}
\label{tab:3678} % is used to refer this table in the text
\end{table*}

\subsection{Implied exomoon mass from the TTVs}

We next turn to the implied exomoon mass, given the TTVs. Unfortunately, with a
TTV signal alone, there is no way to uniquely determine the exomoon period
\citep{kipping:2009a}, which means in turn one cannot invert a TTV amplitude
into an implied mass. Indeed, taking the TTV periodicity and and the planetary
period, and using Equation~(11) of \citet{corridor:2021}, we identify 1260
possible satellite periods that are compatible with the TTV period and lie
between 48.95\% of the Hill sphere and a planet-grazing orbit
($=2\pi\sqrt{R_P^3/(G M_P)}$).

However, this minimum and maximum period range allows to at least evaluate the
range of plausible moon masses needed to explain the TTVs. To rigorously
achieve this though, we first need a posterior distribution for the TTV
amplitude and period. From the 9 available epochs, we decided to drop one epoch
with a partial transit, epoch \#2 on Figure~\ref{fig:gridB}. Partial transits
are challenging to work with but the fact that a large data gap occurs here
potentially compromises our method marginalised detrending scheme. With the 8
remaining epochs, we used \multi\ to regress a sinusoid + linear ephemeris model
and obtain posteriors samples, giving $\Pttv = 2.556_{-0.018}^{+0.018}$\,cycles
and amplitude $A_{\mathrm{TTV}} = 6.95_{-0.33}^{+0.32}$\,minutes.

Equipped with this, we solve for the required satellite mass at each joint
posterior sample in the extreme limits of the widest allowed moon and the
closest allowed moon orbit. Masses are calculated using the
deterministic version of the \forecaster\ code \citep{chen:2017}, as was used
earlier in Section~\ref{sub:ceilingcut}.
In the case of the former, we find that the moon
would need to be just $0.0094_{-0.0031}^{+0.0040}$\,$M_{\oplus}$, or
$0.76_{-0.25}^{+0.33}$\,$M_{\leftmoon}$ - this occurs at a satellite
period of $27.11$\,days. For the closest-in moon, we formally
find $0.34_{-0.11}^{+0.14}$\,$M_{\oplus}$ (occurring at a 3.1\,hour satellite
period) but highlight this is almost certainly an underestimate. For TTVs
periods shorter than the transit duration (10.9\,hours), there is significant
satellite acceleration during the transit that is not correctly accounted for
in the theory of \citet{kipping:2009a}. Accordingly, all we can state with any
confidence is that the putative moon is approximately Lunar mass or greater.
This highlights the significant challenge of confirming such a moon
photometrically, since the dip of such the smallest allowed satellite would
correspond to ${\sim}7$\,ppm. For comparison, the error on the combined
phase-folded Kepler-1513b light curve is $34$\,ppm integrated over the
transit duration timescale, making such a tiny moon wholly undetectable in
direct transit.

\section{Discussion}
\label{sec:discussion}

\subsection{Differences with H+16 work}

In Section~\ref{sub:comparison}, we reported how the H+16 uncertainties are
smaller than that of this work in 66\% of cases, with a mean ratio of a 27\%
difference. The tendency for smaller errors in the H+16 catalog explains why
periodograms from that data set typically return higher significances, leading
to higher peaks in 9-out-of-10 cross-checked cases, as well as why so few of
11 initially identified exomoon corridor candidates survived independent
analysis.

A basic question one might ask is - what explains the differences in transit
times between our own work and that of H+16? There are many distinctions
between the methodologies, any of which could contribute to the differences
in the final product. H+16 use what we would consider to be variant of the
\local\ detrending method, using polynomials but scoring them with $p$-values
from $\mathcal{F}$-tests rather than the BIC, whereas we used a total of eight
detrendings and then method marginalised the ensemble. Transit times were
inferred with an iterative templating procedure in H+16, whereas here a global
light curve model is regressed to the time series. The templating procedure
certainly has the potential to underestimate uncertainties, since the template
is treated as fixed during the transit time search, the error in the transit
shape is not technically propagated into the timing error. Generally, this
might be argued as reasonable since the time of transit minimum has negligible
covariance to the other parameters according to the result of
\citet{carter:2008}. However, that analysis was for a idealised trapezoidal
light curve with no limb darkening, and realistic light curves in general
do exhibit covariances across all parameters, especially when working with
sparse data, partial transits, integration time effects and curved transit
profiles.

In conclusion, we cannot pin-point a single effect that could lead to these
differences as there were simply too many methodology changes, but highlight
the need to be cautious is working with the H+16 TTVs without independent checks.

\subsection{And what of Kepler-1513b?}

The purpose of this paper is to identify interesting exomoon-like TTV signals
from a large ensemble, rather than to perform a detailed analysis of any single
object. As a result, a lot more can and should be done on Kepler-1513b to
uncover its true nature.

We first recommend precise radial velocities to measure the planetary
mass to confirm the planetary nature of the object and better constrain our
moon stability/mass calculations that \forecaster\ can achieve. Such
observations could also identify additional planets in the system that could
potentially explain the TTVs without the need for an exomoon.

Along similar lines, we have only here considered the exomoon hypothesis as a
possible perturber, but a TTV inversion under the planet-planet hypothesis
would reveal the plausibility of such a scenario. Such an inversion is
highly challenging since the period of the putative planet is unknown and
highly multi-modal. Nevertheless, such an analysis could at least identify a
set of plausible alternative periods and interrogate their physical
feasibility.

\section*{Acknowledgments}

% Paul
The authors thank Paul Dalba for graciously sharing his team's spectroscopic parameters ahead of publication.
% Reviewer
The authors would like to thank the anonymous referee for helpful comments that improved this paper.
% Keck
Some of the data used herein were obtained at the W. M. Keck Observatory, which is operated as a scientific partnership among the California Institute of Technology, the University of California, and NASA. The Observatory was made possible by the generous financial support of the W. M. Keck Foundation.
% Hawaii
The authors recognize and acknowledge the cultural role and reverence that the summit of Maunakea has within the indigenous Hawaiian community. We are deeply grateful to have the opportunity to use observations taken from this mountain.
% NASA HEC
Resources supporting this work were provided by the NASA High-End Computing (HEC) Program through the NASA Advanced Supercomputing (NAS) Division at Ames Research Center for the production of the SPOC data products.
% NEA
This research has made use of the NASA Exoplanet Archive, which is operated by the California Institute of Technology, under contract with the National Aeronautics and Space Administration under the Exoplanet Exploration Program.
% Kepler
This paper includes data collected by the Kepler Mission. Funding for the Kepler Mission is provided by the NASA Science Mission directorate.
% Gaia
This work has made use of data from the European Space Agency (ESA) mission Gaia (\href{https://www.cosmos.esa.int/gaia}{https://www.cosmos.esa.int/gaia}), processed by the Gaia Data Process- ing and Analysis Consortium (DPAC, \href{https://www.cosmos.esa.int/web/gaia/dpac/consortium}{https://www.cosmos.esa.int/web/gaia/dpac/consortium}). Funding for the DPAC has been provided by national institutions, in particular the institutions participating in the Gaia Multilateral Agreement.
% Grants
D.K. and D.Y. acknowledge support from NASA Grant \#80NSSC21K0960.
% Donors
This work was enabled thanks to supporters of the Cool Worlds Lab, including
Mark Sloan,
Douglas Daughaday,
Andrew Jones,
Elena West,
Tristan Zajonc,
Chuck Wolfred,
Lasse Skov,
Graeme Benson,
Alex de Vaal,
Mark Elliott,
Stephen Lee,
Zachary Danielson,
Chad Souter,
Marcus Gillette,
Tina Jeffcoat,
Jason Rockett,
Scott Hannum,
Tom Donkin,
Andrew Schoen,
Jacob Black,
Reza Ramezankhani,
Steven Marks,
Gary Canterbury,
Nicholas Gebben,
Joseph Alexander,
Mike Hedlund,
Dhruv Bansal,
Jonathan Sturm,
Rand Corporation,
Leigh Deacon,
Ryan Provost,
Brynjolfur Sigurjonsson,
Benjamin Paul Walford,
Nicholas De Haan,
Joseph Gillmer,
Emerson Garland \&
Alexander Leishman.

\section*{Data Availability}
The code used and results generated by this work are made publcily available
at \wwwcoolworlds.

%\appendix

%

\bsp
\label{lastpage}
\end{document}